\documentclass{iopconfser}
\usepackage{xcolor}
\usepackage{graphicx} % Required for inserting images
\usepackage{mathtools,leftindex,tensor,mhchem,amssymb}
\usepackage{comment}
\usepackage{indentfirst}
\usepackage{float}
\usepackage{url}

\begin{document}

\title{Cluster Reconstruction in Electromagnetic Calorimeters Using Machine Learning Methods}

\author{Kalina Dimitrova$^{1,*}$, Venelin Kozhuharov$^{1}$, Ruslan Nastaev$^{1,**}$ and Peicho Petkov$^{1}$}

\affil{$^1$Faculty of Physics, Sofia University ``St. Kliment Ohridski'', 5 J. Bourchier Blvd., 1164 Sofia, Bulgaria}

\email{$^{*}$ kalina@phys.uni-sofia.bg;\\ ~~~~~~~~~~ $^{**}$ nastaev@uni-sofia.bg}

\begin{abstract}
Machine-learning-based methods can be developed for the reconstruction of clusters in segmented detectors for high energy physics experiments. 
Convolutional neural networks with autoencoder architecture trained on labeled data from a simulated dataset reconstruct events by providing information about the hit point and energy of each particle that has
entered the detector. 
%pp предлагам следващото изречение да стане: 
The correct reconstruction of the position and the energy of the incident particles is crucial for the accurate events reconstruction.
%The correct placement and energy determination 
%are %pp was "is" 
%necessary for the accurate physics events reconstruction. 
The presented method shows the ability to reconstruct the impact point within the same segment as the true position and determines the particle energy with good precision.
%(pp: accuracy -> precision). 
%It can be used in a wide range of cases of data reconstruction where overlapping clusters need to be accurately (трети път... добре е да се замести) distinguished.
It can be applied in a wide range of cases of event reconstruction where the good separation of overlapping signals plays a key role in the data analysis.
\end{abstract}

\section{Introduction}

Over the years, machine learning methods
have emerged as a useful tool in scientific research 
across a wide variety of fields.
In high-energy physics data reconstruction and analysis, 
%plays a crucial role, challenges such as 
pattern recognition and signal processing are necessary steps in the process of obtaining results.
Achieving high performance in identifying individual events and
accurately reading their parameters is possible
only 
%pp
%with 
through
%endpp
the use of reliable algorithms.
%pp
% Modern machine-learning-based techniques 
% such as the
The
%endpp
development of 
convolutional neural networks (CNNs) 
%pp
%enable
enables
%endpp
the training of these algorithms
to analyze experimental data 
by identifying specific patterns.

Pattern reconstruction algorithms need to be developed for experiments 
%pp
%which 
that
%endpp
gather data 
from segmented detectors such as electromagnetic calorimeters.
The waveform 
%pp
of the signal, collected
%/gathered 
%endpp
in each sector of such detectors is generated by the interaction
of a 
%charged particle or a high-energy photon
high energy particle (apart from muons) with the active volume
of the detection material
and the creation of a subsequent
%pp
%.
%endpp
%This interaction provokes a cascade process named 
%electromagnetic 
shower.
The problem with processing 
%these/
such signals
comes from the fact that every individual particle
creates its own shower, which, 
in terms of short-duration sequential events,
can overlap with the neighboring ones
%pp
in both space and time.
%end pp
This may lead to the misinterpretation of the results 
for the properties of the detected particles.
Machine learning methods can provide pattern recognition algorithms, thereby enabling
the separation of overlapping showers~\cite{bib:3Dclu}.

This article introduces the development and application
of two-dimensional CNNs for the reconstruction of clusters in electromagnetic calorimeters.
A dedicated Monte Carlo simulation 
%Kali: да питам Венелин дали да цитирам Монте Карлото на падме или да се правя на ударена
%Peicho: По моето скромно мнение не е добре, човек да се прави на ударен, че може да попадне на рецензент или на читател като Венелин :-P
provides a training dataset that is used for the training of networks with convolutional autoencoder~\cite{bib:autoencoder} architecture. 
The developed method 
%pp
%introduces a modification to 
modifies
%endpp
the classical autoencoder case by using labels in the training,
%pp
instead of supplying the input to the output of the model.
%endpp
%pp
%The
Hence, the 
%endpp
desired output contains information about the positions $(x_i,y_i)$ and energies $E_i$ of the particles.
%the clusters are attributed to.

The analysis of the results compares the real $(x_i,y_i)$ and predicted $(x_i^{pred},y_i^{pred})$ impact positions of the photons from an independent dataset. 
The results for the particle energies after additional post-processing of the predictions are compared to the true values.

For the development and the analysis of the presented methods, 
%pp
we utilised
%endpp
the TensorFlow 
library~\cite{bib:TF} with 
the Keras~\cite{bib:Keras} framework.
The visualisation of the results is done using 
the matplotlib~\cite{bib:matplotlib} library and
the visualkeras~\cite{bib:viskeras} package was used for neural network architecture visualization.
%pp
%The GEANT4 toolkit~\cite{bib:geant} is used in the development of the simulations for generating training and testing datasets. 
The simulations for generating training and testing datasets are based on the GEANT4 toolkit~\cite{bib:geant}.
%endpp

%\section{Simulation of the data and neural network architecture}
\section{Modified autoencoder networks for cluster reconstruction}

%pp
% Due to the nature of the machine learning process, 
% which involves comparing the neural network's
% predictions with true values, 
% using real experimental data
% for training is inefficient.
Because of the nature of the machine learning process, which involves comparing the predictions of the neural network to actual values, training with real experimental data is inefficient.
%endpp
A more effective approach is to employ
a simulation of the experiment,
which provides 
%pp
%clear true
clean ground truth
values 
%and
that
%endpp
can be easily adjusted, if necessary.
%????For the developed neural network, such
%simulations were made.??? -> Does not sound good enough... 
%pp
Hence, the ML models were developed, trained, and validated with simulated data. 
%endpp

\begin{figure}[H]
    \centering
    \includegraphics[width=\linewidth]{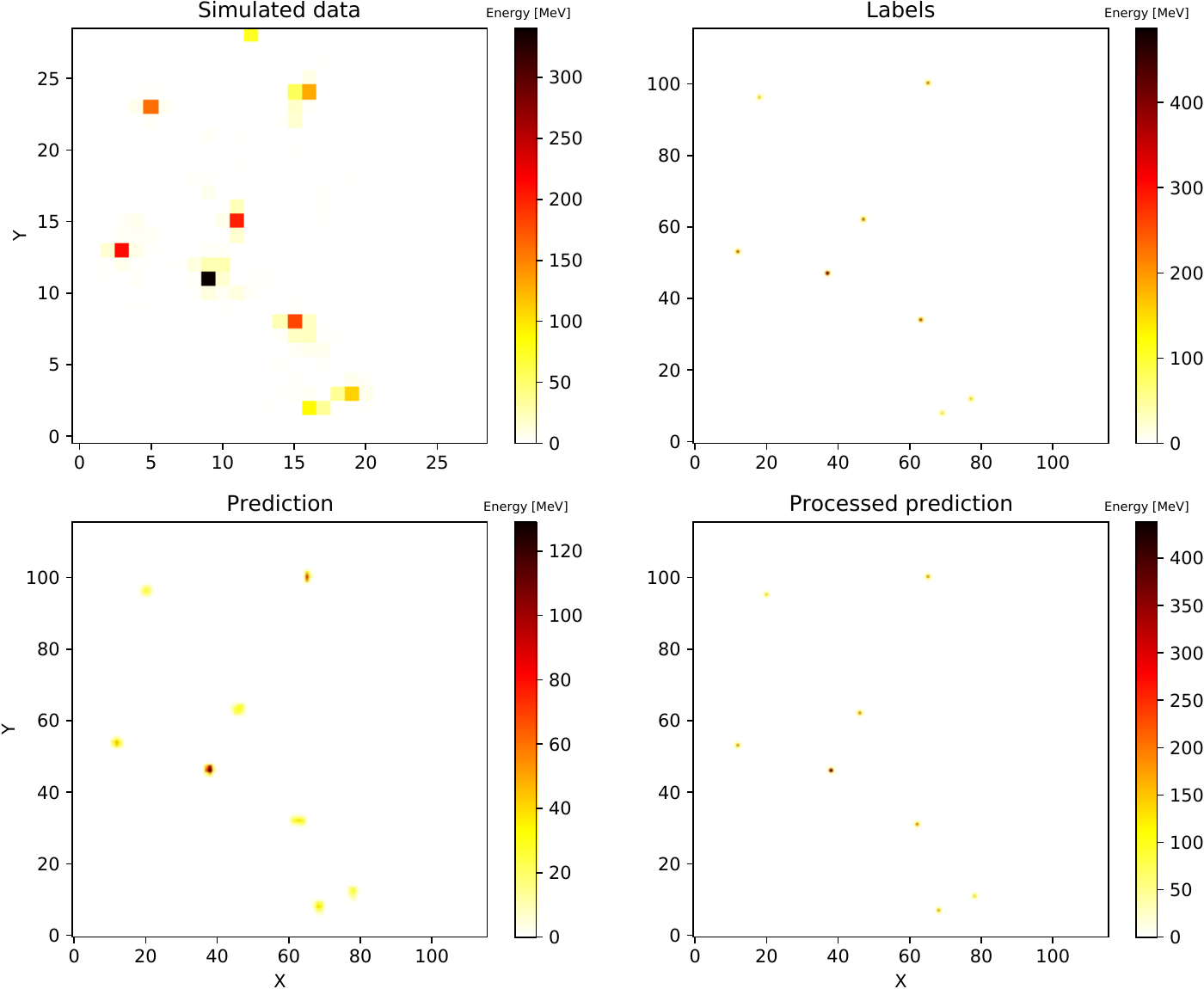}
    \caption{\textbf{Up, left:} A simulated event in a $29\times29$ crystal calorimeter containing several clusters created by photons entering the detector. The value for each crystal represents the energy
      accumulated in the corresponding channel for the whole duration of the event. \textbf{Up, right:} The label, assigned to the event. Each crystal is divided in 16 smaller bins. The ones where a photon has
      hit the detector have the value of the photon's energy, and all others are 0. \textbf{Down, left:} Predicted output by the model. For each recognized cluster, several non-zero values around the impact point are present.
    \textbf{Down, right:} The prediction after post-processing. All values are merged into the maximum one for each recognized cluster, leaving one value for the energy in one position.}
    \label{fig1}
\end{figure}

The simulated electromagnetic calorimeter represents
%is a
a $29\times29$ matrix of scintillating bismuth germanate (BGO) crystals, each with size $2\times2$~cm.
The raw data collected from it represents the accumulated 
%charge 
visible energy
in each crystal due to a particle interacting with it.
%The particles can either be either the initial photons that are produced in physics processes in the experiment or they can belong to the subsequent
%electromagnetic shower that is created due to the photons interacting with the detector.
The developed Monte Carlo simulation produces datasets of events with simulated electromagnetic showers.
%triggered by the hits of photons in the crystals.
%The developed 
A shower triggers several crystals, forming a cluster. 
Each simulated event contains
a certain number of accumulated clusters
%по-скоро са отделните депозирани енергии във всеки кристал... 
caused by photons that have hit the detector during an event duration of 1024~ns. 
The accumulated energy depositions in MeV
%charge in mV 
%deposited energy 
are saved for each crystal, leading to a set of input variables
\begin{equation}
    \epsilon_{k,l}~; ~k ,l \in [1,29],
\end{equation}
where $k$ and $l$ are the row and column in the input matrix.
%\begin{align*}
%k & : \text{row index in the input matrix} \\
%l & : \text{column index in the input matrix} \\
%\end{align*}
The number of 
%clusters 
impinging particles
follows a uniform distribution between 0 and 50, while the energy of the showering particles 
%that created them 
adheres to a Gaussian distribution with a mean value of $\mu = 200$~MeV 
and a standard deviation $\sigma = 200$~MeV.
The training dataset consists of $10^6$ events. An example event from the dataset is shown on the upper left panel of Figure~\ref{fig1}.

%The 
To perform event reconstruction and evaluate  its 
precision 
%of the event reconstruction
%pp
%is directly dependent
%depends directly
%endpp
%on the correct reading of the
%collected data, in this case, the energy carried by the photon and
%the exact position of the impact.
%For this purpose, 
we developed several CNNs
with autoencoder architecture.
All models have similar architectures, consisting of an encoder and a decoder part.
The encoder consists of several 2D convolution layers with a decreasing number of filters with a decreasing kernel size.
Dropout layers are used between the convolution layers.
The decoder features a mirrored structure of the same number of 2D transpose convolution layers, also with dropout layers between them.
The final layer is a transpose convolution layer with a single filter, producing output with a shape resembling the input shape.

%pp
%The training of the models differs from the conventional autoencoder case of unsupervised learning.
Unlike the autoencoder unsupervised learning case, 
%A modification is introduced
we introduced a modification
%endpp
that uses labels of the same shape or an upscaled version as the input data.
The difference between the input and output arrays is that all values in the target output array are set to 0,
except for those that correspond to the crystals where a photon has hit the detector, which are taken to be $E_{k,l}$. 
There, the value represents the
photon's energy. 
In this way, the desired output array contains all the relevant parameters of the event: 
the number of
particles that have arrived in this event, their arrival points, and energies.
A similar approach~\cite{bib:nafski22} is already developed for the case of time series data reconstruction
in individual crystals with the goal of reconstructing the precise arrival times and amplitudes
of the pulses created by particles entering the crystal. 
In this notations the role of the model is to perform the mapping 
\begin{equation}
    \epsilon_{k,l} ~ \longrightarrow E_{k,l}^{pred},
\end{equation}
and the minimization is done over
\begin{equation}
    \chi^2 = \sum_{k,l} (E_{k,l} - E_{k,l}^{pred})^2.
\end{equation}

Upscaling of the output is applied for a more precise reconstruction of the particle's impact point and increasing the spatial resolution.
To achieve this, each crystal is divided into 16 smaller square bins measuring $0.5\times0.5$~cm. The particle energy is placed in the bin, corresponding to its arrival position, and all others are set to 0. Each $29\times29$ input event is therefore assigned a 4 times upsampled label of $116\times116$ values.
An example label is shown on the upper right panel of Figure~\ref{fig1}. To ensure the correct output shape, two $2\times2$ 2D upsampling
layers are introduced in the model architectures between the decoder's transpose convolution layers.

An example prediction is shown on the lower left panel of Figure~\ref{fig1}. 
For each recognized cluster,  several non-zero
values around the maximum one
 can be seen.
The coordinates $(x_{kl},y_{kl})$ of the cell with the maximal $E_{K,L}^{pred}$
%which 
would be the predicted impact point. 
To get the full value $E_i^{pred}$ of the predicted energy
of a cluster with index $i$
an additional post-processing algorithm performs convolution across the 
predicted output array by moving a
$5\times5$ window across it and adding all values in this window into the maximum one:
\begin{equation}
    E_i^{pred} = \sum_{n=-2}^{2} \sum_{m=-2}^2 E_{K+n,L+m}^{pred}.
\end{equation}
After a value $E_{K+n,L+m}$ is added to an identified cluster, 
it is set to zero and the cluster identification procedure continues
with the next local maximum in the $5\times5$ window. 

An example of a post-processed prediction is
shown on the lower right panel of Figure~\ref{fig1}.

 %Kali: формула за пост обработката?

\section{Results}

\begin{figure}[H]
    \centering
    \includegraphics[width=0.47\linewidth]{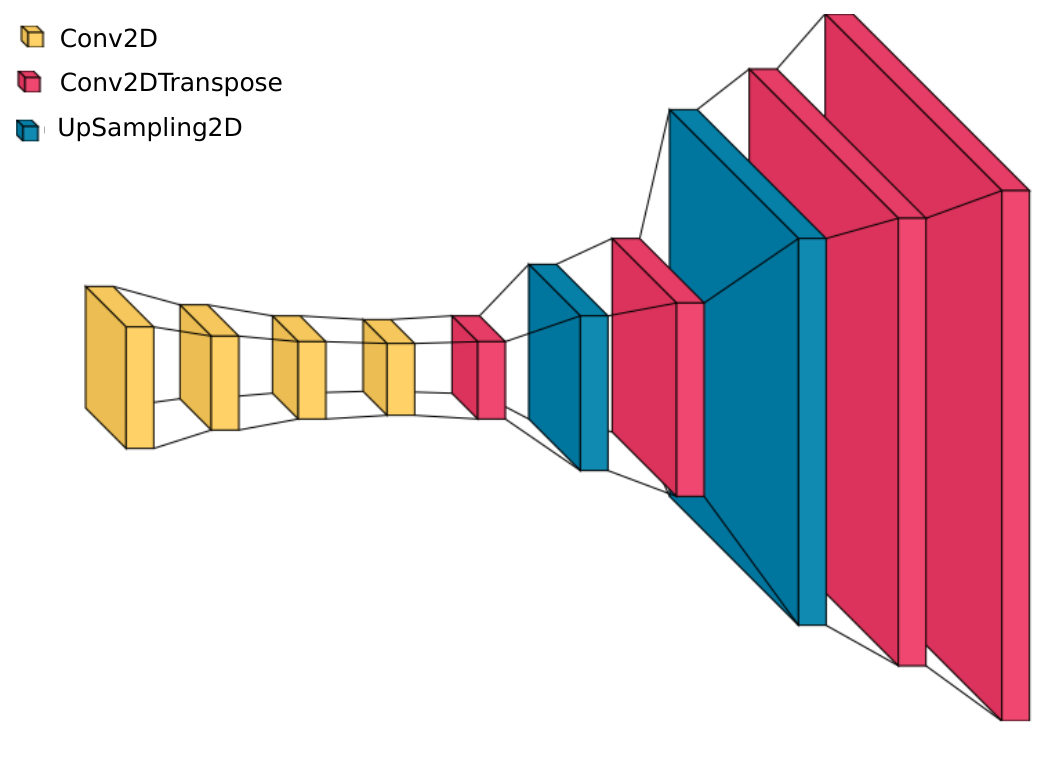}
    \includegraphics[width=0.48\linewidth]{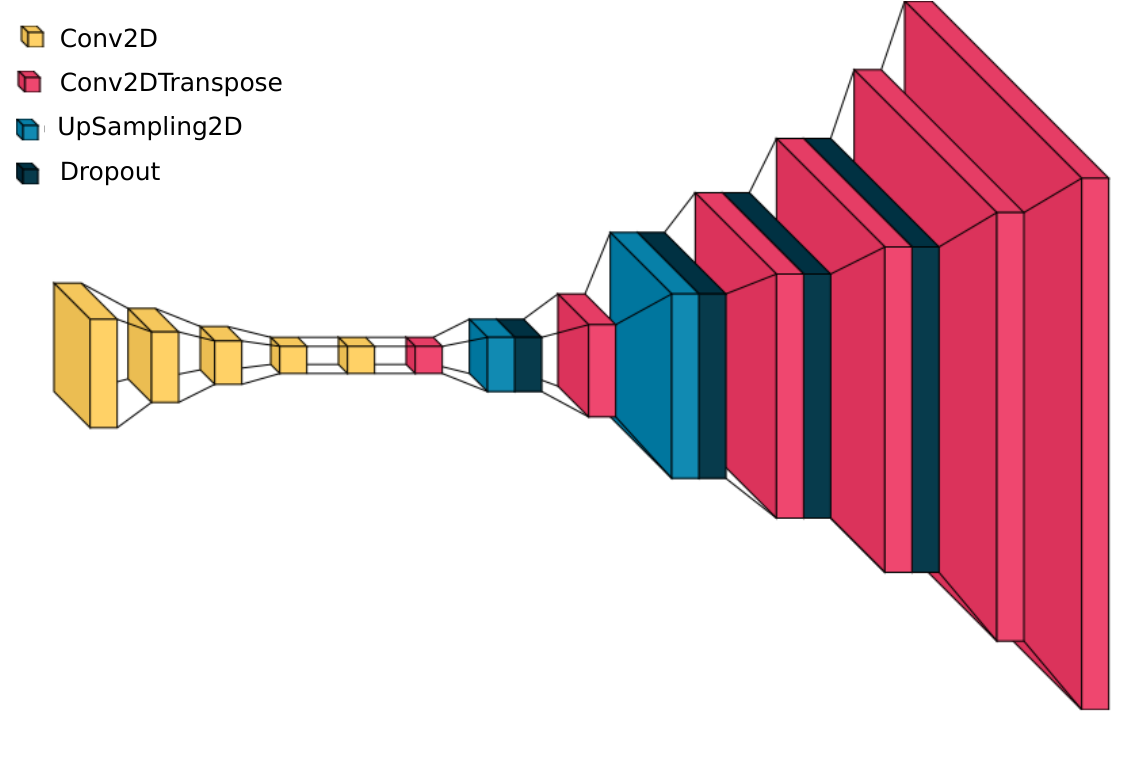}
    \caption{Structure of the hidden layers in the two tested models. Both models consist of an encoder and decoder part with upsampling layers added in the latter. \textbf{Left:} Model 1 has an encoder or 4 convolutional layers (yellow) and a decoder of 4 transposed convolution layers (pink). A final transposed convolution layer with 1 filter provides the output. \textbf{Right:} Model 2 has deeper architecture with 5 convolution layers in the encoder and decoder and dropout layers (dark blue) in the decoder.}
    \label{fig:arch}
\end{figure}

Several models with a different number of layers and dropout parameters were developed and tested.
Model 1 has 4 layers in the encoder and 4 layers in the decoder, with no dropout layers between them, while Model 2 has a 5-layer encoder and decoder and dropout layers
with a 0.2 rate in the decoder. The architectures of the two models are shown on Figure~\ref{fig:arch}.
They are applied to an independent test dataset, and the predictions are compared to the ground truth labels.

The first important goal of any clusterization algorithm is to correctly identify as many clusters as possible without the generation of noise.
In order to find out which are recognized, a check is performed for all simulated clusters. For any given cluster from the output array of the event, the predictions array is checked for
non-zero values in a
$5\times5$ window around the simulated interaction point. If a prediction is found within the window, it's associated with that cluster, and a match is declared.

\begin{figure}[h]
    \centering
    \includegraphics[width=0.49\linewidth]{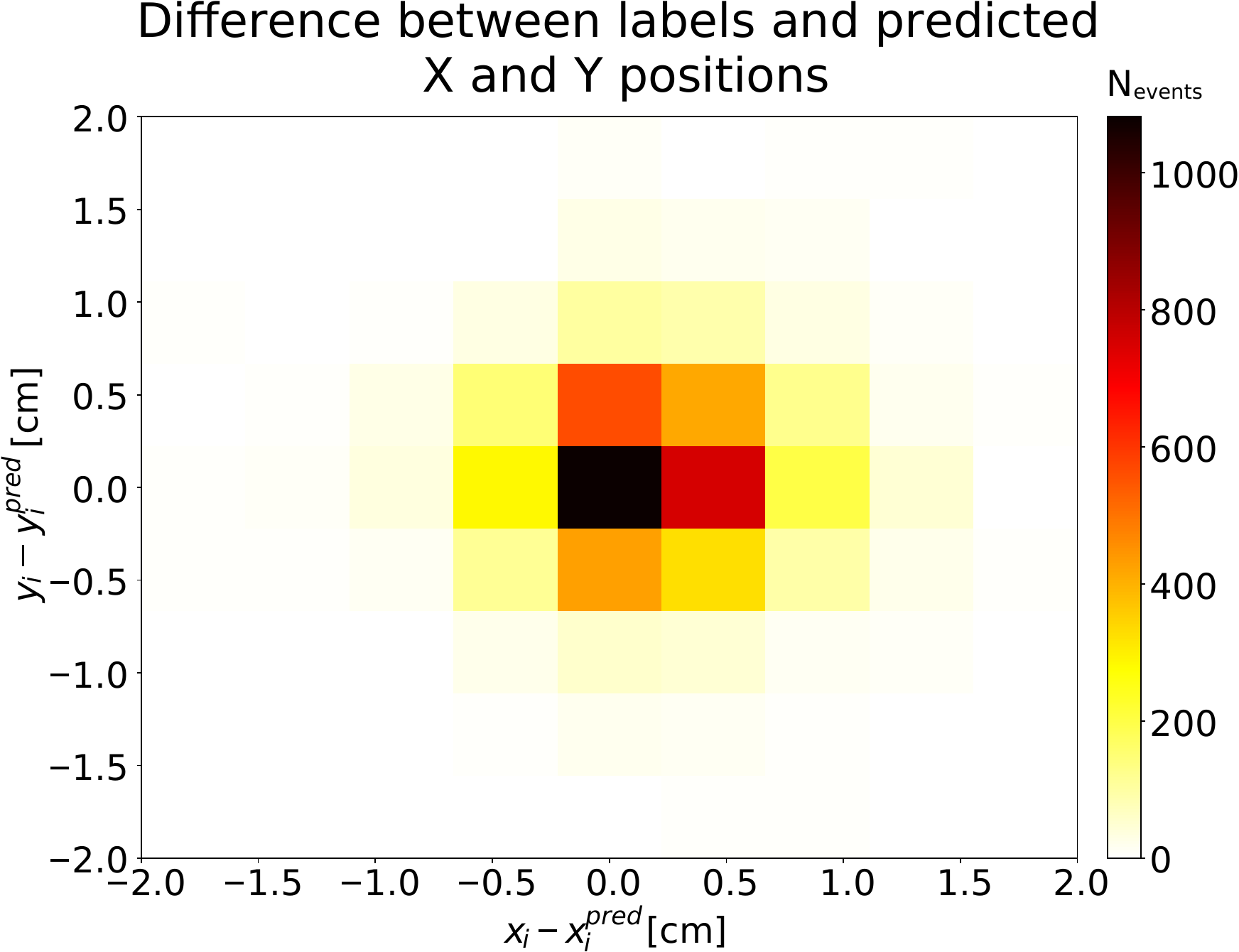}
    \includegraphics[width=0.48\linewidth]{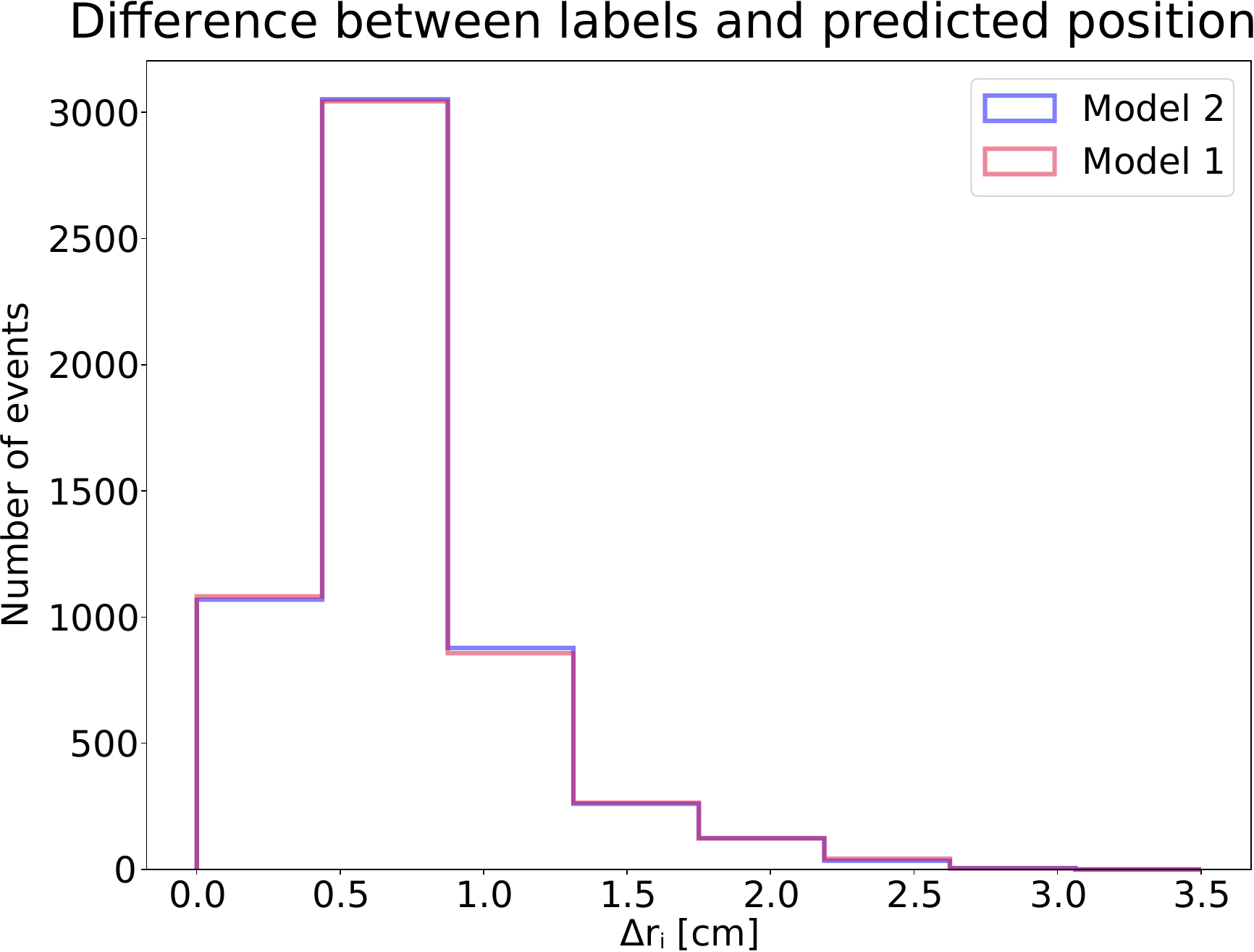}
    \caption{Difference between the true and the predicted position for all found clusters in the test dataset.  \textbf{Left:} Distribution of the individual offsets of the found clusters. Part of the events have $\Delta x_i>0$, which means they are placed at a smaller $x$ than their true position.
    \textbf{Right:}. Distribution of the distance between the reconstructed position and the label for all found clusters. A big number of events are predicted less than 1~cm away from the true position of the hit. Events at larger offsets might be mismatched predictions.}
    \label{fig2}
\end{figure}

To evaluate the accuracy of the hit position reconstruction, the difference between the actual and reconstructed position 
\begin{equation}
    \Delta r_i =  \sqrt{(x_i - x_i^{pred})^2 + (y_i - y_i^{pred})^2 }
\end{equation}
is calculated for all recognized clusters. Figure~\ref{fig2} shows the distribution of this
offset independently for the two axes $\Delta x_i = x_i - x_i^{pred}$ and $\Delta y_i = y_i - y_i^{pred}$, as well as combined as the distance $\Delta r_i$.  The two models have almost identical performance in determining the position, with both reconstructing a large number of the events within less than 1~cm from the actual hit point. The distribution of $\Delta x_i$ has a mean value 
$\overline{\Delta_1 x} = 0.199$~cm and $\sigma_1(\Delta x)=0.535$~cm for model 1 and $\overline{\Delta_2 x} =  0.223$~cm and $\sigma_2(\Delta x)= 0.537$~cm for model 2. $\Delta y_i$ has a mean value 
$\overline{\Delta_1 y}=0.054$~cm and $\sigma_1(\Delta y)=0.533$~cm for model 1 and $\overline{\Delta_2 y}=0.062$ and $\sigma_2(\Delta y)=0.518$ for model 2.
Both models show a 
small bias  (of the order of 2 mm, half a cell)
%small deviation 
when reconstructing the $x$ coordinate of the clusters, with the reconstructed cluster being placed at a smaller $x$ than the position in the label.

\begin{figure}[H]
    \centering
    \includegraphics[width=0.9\linewidth]{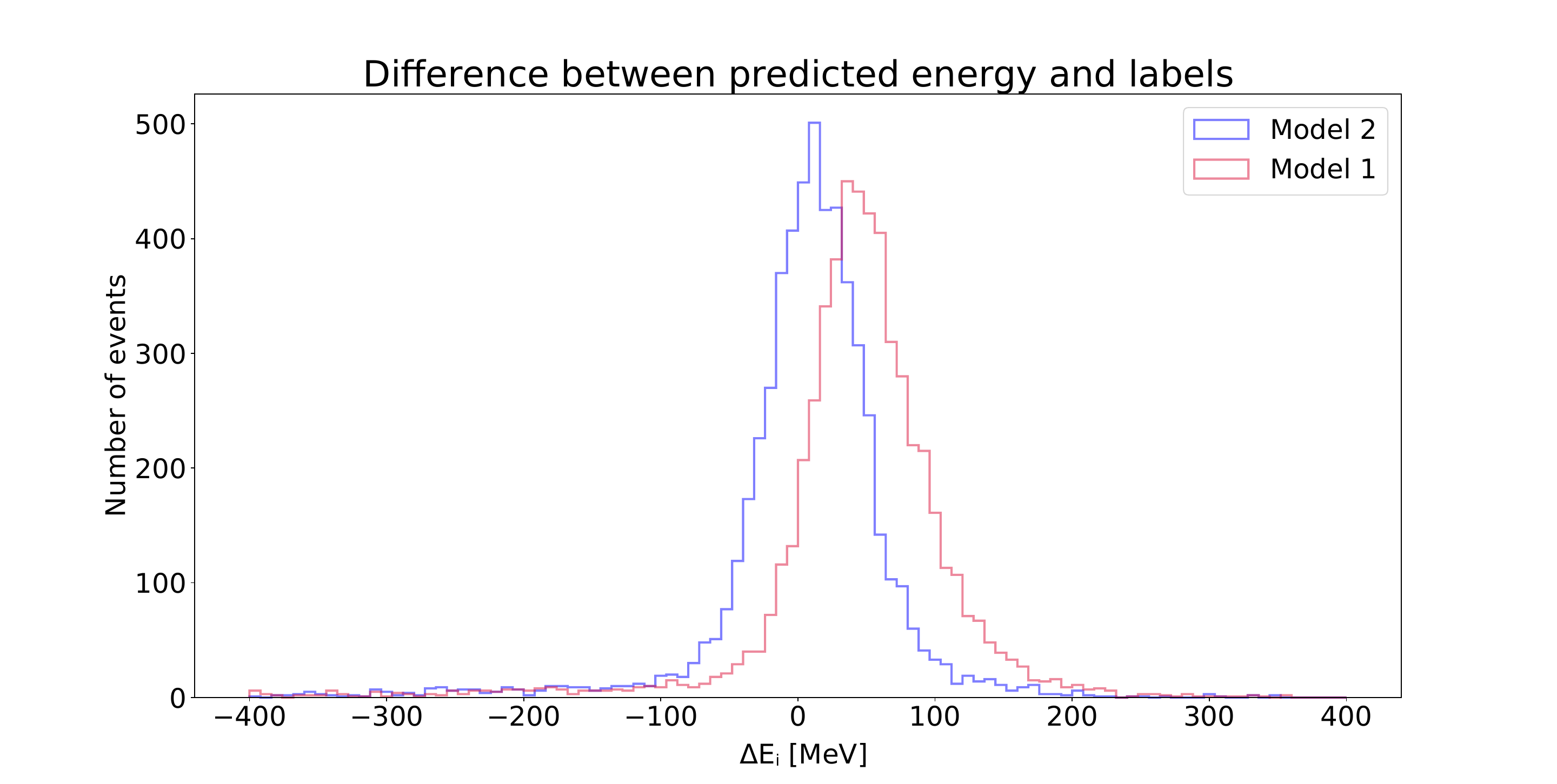}
    \caption{Difference between the true and the predicted values for the energy of the particles the identified clusters originate from. Model 2 shows better performance, having the distribution centered at 0.}
    \label{fig3}
\end{figure}

Figure~\ref{fig3} shows
the difference between the true $E_i$ and the predicted $E_i^{pred}$ energy, 
\begin{equation}
    \Delta E_i = E_i - E_i^{pred}
\end{equation}
for the identified clusters, 
using the two different models. 
Both models have similar width of the distribution, with $\sigma_1(\Delta E) = 74.83~ \mathrm{MeV}$, and $\sigma_2(\Delta E) = 75.46~ \mathrm{MeV}$. However, the mean value for model 2 is $\overline{\Delta_2 E}  = 3.15~ \mathrm{MeV}$  and is close to zero, compared to $\overline{\Delta_1  E}  = 42.39~ \mathrm{MeV}$. This indicates that model 2 which has a deeper architecture shows much better performance for the energy reconstruction.

The observed identified and reconstructed clusters  with very big 
$\Delta E$  might again be due to a mismatch between a true cluster and a predicted one.
%Напиши средните стойности и сигма/RMS на разпределенията. Тогава лесно се сравняват

\section{Conclusions and discussion}

In the reconstruction and subsequent analysis of data from electromagnetic calorimeters,
achieving high performance in
identifying individual events
and accurately reading their parameters
directly depends on effectively
addressing challenges in
pattern recognition and signal processing.
Machine learning methods show promising results in that direction. Several models for cluster reconstruction with autoencoder architecture were developed and tested,
using Monte Carlo simulations of a
segmented scintillating crystal detector.
%a detector composed of BGO crystals. 
The networks are trained on labeled data, and their performance is further enhanced by introducing
upsampling layers in the model architecture.

The results for two models with different depths are compared.
The depth of the model has little influence over the position reconstruction.
Most of the predicted clusters
are placed less than 1~cm away from
the actual hit point, which gives precision of less than 
half the chosen crystal size.
%pp
%The reconstructed particle energy shows a good relation to the true one.
The reconstructed particle energy is in good agreement with the true one.
%endpp

The promising results of
the developed model demonstrate the
potential of the chosen approach.
Higher precision could be achieved
by increasing the training dataset size
along with further modifications
to the model’s architecture and testing deeper models.
One of the main problems the model was designed to solve is the need for 
separation of closely placed clusters. Further analysis of the results will provide
insight into the cluster separation abilities of the method.

\section*{Acknowledgements}
This work is partially supported by BNSF: KP-06-D002\_4/15.12.2020 within MUCCA, CHIST-ERA-19-XAI-009 and by the European Union - NextGenerationEU, through the
National Recovery and Resilience Plan of the Republic of Bulgaria, project SUMMIT BG-RRP-2.004-0008-C01.

\end{document}